\documentclass[aps,prd,showpacs,nofootinbib,floats,floatfix,amssymb,amsmath,
preprintnumbers,twocolumn]{revtex4}

\usepackage{bm}
\usepackage{latexsym}
\usepackage{dcolumn}
\usepackage{amsfonts,amssymb}
\usepackage{graphicx,epsfig}
\usepackage{psfrag}
\usepackage{amsmath,amssymb}

\def\beq{\begin{equation}}
\def\eeq{\end{equation}}
\def\br{\begin{eqnarray}}
\def\er{\end{eqnarray}}
\def\benu{\begin{enumerate}}
\def\eenu{\end{enumerate}}
\def\nn{\nonumber} 

\def\l{\left}
\def\r{\right}

\newcommand{\pbox}[1]   {%
\makebox[0pt][r]{\raisebox{7mm}[0pt][0pt]{\footnotesize #1}}\ignorespaces}
\newcommand{\be} {\begin{equation}}
\newcommand{\ee}[1] {\label{#1}\end{equation}\pbox{#1}}

\begin{document}

\title{Path integral duality and Planck scale corrections to the\\
primordial spectrum in exponential inflation}
\author{L.~Sriramkumar}\email[]{E-mail: sriram@mri.ernet.in}
\affiliation{Harish-Chandra Research Institute, Chhatnag Road, Jhunsi,
Allahabad~211~019, India.}
\author{S.~Shankaranarayanan}\email[]{E-mail: shanki@aei.mpg.de}
\affiliation{HEP Group, International Centre for Theoretical Physics,
Strada costiera~11, 34100~Trieste, Italy \\
Max-Planck-Institut f\"ur Gravitationsphysik, 
Albert-Einstein-Institut, 
Am M\"uhlenberg 1, D-14476 Potsdam, Germany}\date{\today}


\begin{abstract}
The enormous red-shifting of the modes during the inflationary epoch 
suggests that physics at the Planck scale may modify the standard, 
nearly, scale-invariant, primordial, density perturbation spectrum.
Under the principle of path-integral duality, the space-time behaves 
as though it has a minimal length~$L_{_{\rm P}}$ (which we shall 
assume to be of the order of the Planck length), a feature that is 
expected to arise when the quantum gravitational effects on the matter 
fields have been taken into account.
Using the method of path integral duality, in this work, we evaluate 
the Planck scale corrections to the spectrum of density perturbations 
in the case of exponential inflation.
We find that the amplitude of the corrections is of the order of 
$({\cal H}/M_{_{\rm P}})$, where ${\cal H}$ and $M_{_{\rm P}}$ denote 
the inflationary and the Planck energy scales, respectively.
We also find that the corrections turn out to be completely 
independent of scale.
We briefly discuss the implications of our result, and also comment 
on how it compares with an earlier result. 
\end{abstract}
\maketitle


\section{Introduction}\label{intro}

The inflationary scenario~\cite{texts,mukhanov92} provides an
attractive mechanism to generate the primordial density perturbations
that are eventually responsible for the anisotropies in the cosmic
microwave background~(CMB) and the formation of the large scale
structure~(LSS).  Typically, the CMB and LSS data have been used to
constrain the parameters of the inflationary model such as, for
example, the height and the slope of the canonical scalar field
potential~\cite{lidsey97,bassett05}.  However, over the last few
years, it has been realized that, in most of the models of
inflation~\cite{lyth99}, the period of acceleration lasts so long that
length scales that are of cosmological interest today would have
emerged from sub-Planckian length scales at the beginning of
inflation.  In other words, inflation provides a magnifying glass to
probe physics at the very high energy scales (say, $\sim 10^{17}\,
{\rm GeV}$) that are otherwise inaccessible to observations or
experiments.  This has led to a considerable effort in the literature
towards understanding the effects of Planck scale physics on the
inflationary perturbation spectrum~\cite{paddyolp}--\cite{sriram05},
and the resulting signatures on the
CMB~\cite{minimalcmb,nccmb,lqgcmb,generalcmb}.

In an exponential (i.e. de Sitter) or a power-law inflationary
scenario, the spectrum of perturbations is essentially given by the
Fourier transform of the Wightman function of a free, massless and
minimally coupled scalar field (see, for instance,
Ref.~\cite{mukhanov92}).  Therefore, in order to study the effects of
Planck scale physics on the primordial spectrum, we need to understand
as to how quantum gravitational effects modify the two-point function
of a scalar field in an inflationary background.  But, due to the lack
of a viable quantum theory of gravitation, one is forced to consider
phenomenological models constructed by hand---models which are
supposed to contain one or more features of the actual effective
theory obtained by integrating out the gravitational degrees of
freedom.  The different high energy models that have been considered
in this context modify either the dynamics or the initial conditions
(and, in some cases, both) of the canonical scalar
field~\cite{modifieddr}--\cite{sriram05}.

A class of models that has been used extensively to study the Planck
scale corrections to the primordial spectrum involves the violation of
local Lorentz invariance at the energy scales of
inflation~\cite{modifieddr,shanki05}.  However, theoretically, there
exists no apriori reason to believe that Lorentz invariance may be
broken at the inflationary energy scale.  Moreover, certain
observations and experiments seem to indicate that local Lorentz
invariance may be preserved to extremely high energies (see
Ref.~\cite{vlorentzi}; in this context, also see
Ref.~\cite{vlorentzic}).  In such a situation, to study the Planck
scale effects, it becomes important to consider models that {\it
preserve}\/ local Lorentz invariance {\it even as they contain a
fundamental scale}\/~\cite{shanki04}.

One such model that introduces a fundamental length scale while
preserving Lorentz invariance is the approach due to the principle of
path-integral duality~\cite{paddypid}. Recall that, in standard
quantum field theory, the path integral amplitude for a path
connecting events~${\tilde x}$ and~${\tilde x'}$ in a given space-time
is proportional to the proper length, say, ${\cal R}({\tilde
x},{\tilde x'})$ between the two events. The duality principle
proposes that the path integral amplitude should be invariant under
the transformation ${\cal R} \rightarrow \l(L_{_{\rm P}}^2/{\cal
R}\r)$, where $L_{_{\rm P}}$ is assumed to be of the order of the
Planck length.  A consequence of this postulate is that, in the
two-point function in the Minkowski vacuum, the proper distance, say,
$(\Delta {\tilde x})^2$ between two space-time events is replaced
by~$\left[(\Delta {\tilde x})^2 + 4\, L_{_{\rm P}}^2\right]$. This
suggests that, under the duality principle, the background space-time
behaves as though it possesses a minimal length~$L_{_{\rm P}}$, a
feature that is expected to arise when the quantum gravitational
effects are taken into account.

In this work, we shall utilize the locally Lorentz invariant approach
due to path integral duality to evaluate the Planck scale corrections
to the spectrum of density perturbations in the exponential
inflationary scenario.  It should be pointed out here that, recently,
another approach that is based on T-duality in string theory has been
used to evaluate the Planck scale corrections to the standard spectrum
of gravitational waves in exponential inflation~\cite{chouha05}.
There seem to exist some correspondence between the approaches of
T-duality in string theory and the principle of path integral
duality\footnote{For instance, using the T-duality in string theory,
it has been shown that the propagator for the string center of mass
results in the same propagator as obtained through the principle of
path integral duality. For details, see Ref.~\cite{Fontanini}.}.
However, we would like to emphasize here the following crucial
difference between the approach motivated by T-duality in string
theory that has been adopted in Ref.~\cite{chouha05} and the approach
due to path integral duality we shall adopt.  In Ref.~\cite{chouha05},
the authors use a modified dispersion relation to mimic the effects
due to T-duality on the propagator.  Clearly, such an approach assumes
violation of local Lorentz invariance.  In this work, we shall instead
use the proper time representation of the modified
propagator~\cite{paddypid}---an approach that explicitly preserves
local Lorentz invariance---to evaluate the Planck scale corrections to
the spectrum of perturbations.  As we shall see, the amplitude of the
corrections prove to be the order of $({\cal H}/M_{_{\rm P}})$, where
${\cal H}$ and $M_{_{\rm P}}$ denote the inflationary and the Planck
energy scales.  Moreover, in contrast to the results obtained in
Ref.~\cite{chouha05}, we find that the corrections turn out to be
completely independent of scale.

The remainder of this paper is organized as follows.  The
modifications to the two-point functions that arise due to the path
integral duality are easily expressed in terms of the Schwinger's
proper time representation for the Greens functions~\cite{paddypid}.
Therefore, in the following section, using the proper time method, we
shall very briefly outline as to how the approach of path integral
duality modifies the two-point functions of a quantum scalar field
propagating in a curved space-time.  We shall also point out the
initial conditions that need to be imposed for evaluating the
perturbation spectrum through the proper time method.  In
Section~\ref{sec:psptm}, we shall obtain the standard, scale invariant
perturbation spectrum in exponential inflation through the proper time
formalism.  In Section~\ref{sec:corrections}, we shall evaluate the
Planck scale corrections to the scale invariant spectrum using the
approach of path integral duality.  Finally, in
Section~\ref{sec:discussion}, we conclude with a summary and
discussion of the results we have obtained.

A few words on our convention and notation are in order at this stage
of our discussion.  The metric signature we shall adopt is $(+, -,-,
-)$, and we shall set $\hbar=c=1$.  Also, for convenience, we shall
denote the set of four coordinates $x^{\mu}$ as ${\tilde x}$.

\section{Path integral duality modified two-point 
functions}\label{sec:pid}
 
Consider a free and minimally coupled scalar field of mass $m$ that is
propagating in a classical gravitational background described by the
metric tensor $g_{\mu\nu}$.  In Schwinger's proper time formalism, the
Feynman Green's function corresponding to such a scalar field can be
expressed as~\cite{schwinger51,dewitt75}
\beq
G_{\rm F}({\tilde x},{\tilde x'})
=i\int\limits_0^{\infty}ds\; e^{-i m^2 s}\;
K({\tilde x}, {\tilde x'}; s),
\label{eq:FeyGree-def}
\eeq
where $K({\tilde x}, {\tilde x'}; s)$ is defined as
\beq 
K({\tilde x}, {\tilde x'}; s) \equiv \langle {\tilde x}\vert 
e^{-i {\widehat \Box} s} \vert {\tilde x'}\rangle.  
\label{eq:propti-K}
\eeq
In other words, the quantity $K({\tilde x}, {\tilde x'}; s)$ is the 
path integral amplitude for a quantum mechanical system described by 
the following Hamiltonian:
\beq
{\widehat {\sf H}} = {\widehat \Box} 
\equiv \frac{1}{\sqrt{-g}} 
\partial_{\mu}\l(g^{\mu\nu} \partial_{\nu} \r).
\label{eq:Ham-KG-def}
\eeq

It can be shown that demanding the principle of path integral 
duality corresponds to modifying the above expression for the 
Feynman propagator to (for details, see Refs.~\cite{paddypid})
\beq
G^{\rm (M)}_{\rm F}({\tilde x},{\tilde x'})
=i\int\limits_0^{\infty}ds\; e^{i L_{_{\rm P}}^2/s}\;
e^{-i m^2 s}\; K({\tilde x}, {\tilde x'}; s),\label{eq:FG-Mod}
\eeq
where, as pointed out earlier, $L_{_{\rm P}}$ is of the order of the
Planck length.  We shall utilize this prescription to evaluate the
Planck scale corrections to the inflationary perturbation spectrum.
However, before proceeding with the calculations, we need to clarify
two technical issues.

Firstly, in cosmological perturbation theory, the spectra of
perturbations are determined by the Fourier transform of the Hadamard
function (i.e. the symmetric two-point function) of the quantum field.
The two-point function~(\ref{eq:FeyGree-def}), however, is a Feynman
Green's function rather than the Hadamard function.  It can be shown
that, for a minimally coupled scalar field, the Hadamard function can
be written as\footnote{See Appendix~\ref{app:hadamard}. In this
appendix, we explicitly illustrate this result for the case of the
Hadamard function in the Minkowski vacuum.}
\beq
G^{(1)}({\tilde x},{\tilde x'})
=\int\limits_{-\infty}^{\infty}ds\; e^{-im^2 s}\;
K({\tilde x}, {\tilde x'}; s),
\label{eq:defHada}
\eeq
where $K({\tilde x}, {\tilde x'}; s)$ is the kernel defined in
Eq.~(\ref{eq:propti-K}).  Therefore, the path integral duality
modified Hadamard function can be written as
\beq
G_{\rm (M)}^{(1)}({\tilde x},{\tilde x'})
=\int\limits_{-\infty}^{\infty}ds\; e^{i L_{_{\rm P}}^2/s}\;
e^{-im^2 s}\; K({\tilde x}, {\tilde x'}; s).
\label{eq:mHada}
\eeq

Secondly, in a time-dependent background, since the field evolves with
time, the states at two different instants are, in general, different.
In the inflationary scenario, the perturbations are evaluated as
expectation values in the in-vacuum, a state which is defined when the
modes are well inside the Hubble radius (see our discussion
immediately before and after Eq.~(\ref{eq:psdfntn}) below).
Therefore, when evaluating the kernel $K({\tilde x}, {\tilde x'}; s)$,
we need to impose suitable initial conditions to ensure that the
resulting Green function is an in-in function (rather than the in-out
function that is usually considered in, say, S-matrix or scattering
calculations).

In the next section, we shall obtain the scale invariant perturbation
spectrum in a de Sitter universe using the
expression~(\ref{eq:defHada}) for the Hadamard function.  And, in
Section~\ref{sec:corrections}, we shall evaluate the Planck scale
corrections to the scale invariant spectrum using the modified
Hadamard function~(\ref{eq:mHada}).

\section{Perturbation spectrum through the proper time 
method}\label{sec:psptm}

As we had mentioned in the introduction, in an exponential
inflationary scenario, the metric perturbations (both scalar and
tensor perturbations) can be described by a massless and minimally
coupled scalar field, say, $\Phi$, which satisfies the following
equation of motion~\cite{texts,mukhanov92}:
\beq
\Box\Phi=0.
\eeq
Also, the perturbations are assumed to be induced by the fluctuations
in the free quantum field ${\hat\Phi}$.  Therefore, the power spectrum
as well as the statistical properties of the perturbations are
completely characterized by the two-point functions of the quantum
field.  The power spectrum of the scalar perturbations per logarithmic
interval, viz. $\left[k^3\, {\cal P}_{\Phi}(k)\right]$, is defined
as~\cite{texts,mukhanov92}
\br
\!\!
\int\limits_{0}^{\infty}\! \l(\frac{dk}{k}\r)\, 
\left[k^3\; {\cal P}_{\Phi}(k)\right] 
\!\!&=&\!\!\langle 0 \vert {\hat \Phi}^2 (\eta, {\bf x})
\vert 0\rangle\nn\\ 
\!\!&=&\!\! G^{+}\l({\tilde x},{\tilde x}\r)
=\l(\frac{1}{2}\r)G^{(1)}\l({\tilde x},{\tilde x}\r),\quad
\label{eq:psdfntn}
\er
where $\vert 0\rangle$ is the vacuum state of the field, and
$G^{+}({\tilde x}, {\tilde x'})$ and $G^{(1)}({\tilde x}, {\tilde
x'})$ denote the Wightman and the Hadamard functions of the quantum
field, respectively.

Consider a spatially-flat Friedmann universe described by the 
line-element
\beq
ds^2=dt^2-a^{2}(t)\, d{\bf x}^2,\label{eq:frwle}
\eeq
where $t$ is the cosmic time and $a(t)$ denotes the scale factor.  The
symmetry of the Friedmann background allows us to write the kernel
$K({\tilde x},{\tilde x'};s)$ as follows (see, for example,
Ref.~\cite{duru86}):
\beq
K({\tilde x},{\tilde x'};s)
=\int\!\frac{d^3{\bf k}}{(2\pi)^3}\,
e^{i{\bf k}\cdot\l({\bf x}-{\bf x'}\r)}\, 
\langle t\vert e^{i\, {\hat {\sf H}}_k\, s} \vert t'\rangle,
\label{eq:kernelfrw}
\eeq
where 
\beq
{\hat {\sf H}}_{k}\equiv 
-\l(\frac{1}{a^3}\r)\frac{d}{dt}\l(a^3\frac{d}{dt}\r)
-\l(\frac{k^2}{a^2}\r).\label{eq:defHk}
\eeq
Therefore, for a massless scalar field, the Hadamard
function~(\ref{eq:defHada})  is given by
\beq
G^{(1)}({\tilde x},{\tilde x'})
=\int\limits_{-\infty}^{\infty}\!ds\,
\int\!\frac{d^3{\bf k}}{(2\pi)^3}\,
e^{i{\bf k}\cdot\l({\bf x}-{\bf x'}\r)}\, 
\langle t\vert e^{i\, {\hat {\sf H}}_k\, s} \vert t'\rangle.
\label{eq:gfns}
\eeq
From this expression and the definition~(\ref{eq:psdfntn}) 
of the power spectrum, we then obtain 
\br
\left[k^3\; {\cal P}_{\Phi}(k)\right] 
&=&\l(\frac{k^{3}}{4\, \pi^{2}}\r)
\int\limits_{-\infty}^{\infty}\! ds\; 
\langle t\vert e^{i\, {\hat {\sf H}}_k\, s} \vert t\rangle\nn\\
&=&\l(\frac{k^{3}}{4\, \pi^{2}}\r)
\int\limits_{-\infty}^{\infty}\! ds\; 
\langle t\vert e^{-i\, {\hat {\sf H}}_k\, (-s)} \vert t\rangle.
\label{eq:PS1}
\er

The Schrodinger equation corresponding to the Hamiltonian 
${\hat {\sf H}}_{k}$ is given by
\beq
{\hat {\sf H}}_{k}\, \psi_{E}
\equiv -\l(\frac{1}{a^3}\r)\frac{d}{dt}\l(a^3\frac{d\psi_{E}}{dt}\r)
-\l(\frac{k^2}{a^2}\r)\psi_{E}=E\, \psi_{E},
\label{eq:sch1}
\eeq
where the wave function $\psi_{E}$ is normalized as follows:
\beq
\int\limits_{{\rm all}\; t}\! dt\, a^{3}(t)\, \psi_{E}(t)\,
\psi_{E'}^{*}(t)=\delta\l(E-E'\r).\label{eq:norm1}
\eeq
(In the de Sitter background that we are interested in, the energy $E$
turns out to be a continuous distribution. Hence we have normalized
the wave functions using the delta function normalization.)  If we
write
\beq
\psi_{E}(t)=a^{-3/2}(t)\; \chi_{E}(t),\label{eq:sub1}
\eeq
then we find that the function $\chi_{E}(t)$ satisfies the 
differential equation
\br
& &\!\!\!\!\!\!\!\!\!\!\!\!\!\!\!\!\!\!
-\l(\!\frac{d^2 \chi_{E}}{dt^2}\!\r)
-\l[\l(\frac{k^2}{a^2}\!\r)-\l(\frac{3}{4}\r)\! 
\l(\frac{\dot a}{a}\r)^2-\l(\frac{3}{2}\r)\! 
\l(\frac{\ddot a}{a}\r)\r]\chi_{E}\nn\\
& &\qquad\qquad\qquad\qquad\qquad\qquad\qquad\quad\;
=E\, \chi_{E},\label{eq:sch2}
\er
and the normalization condition~(\ref{eq:norm1}) reduces to
\beq
\int\limits_{{\rm all}\; t}\! dt\, \chi_{E}(t)\,
\chi_{E'}^{*}(t)=\delta\l(E-E'\r).
\label{eq:norm2}
\eeq

On using the Feynman-Kac formula (see, for instance, 
Refs.~\cite{fh65,schulman81}), we can write the kernel 
$\langle t\vert e^{-i\, {\hat H}_k\, (-s)} \vert t\rangle$ 
in terms of the wave function $\chi_{E}$ as follows:
\br
\langle t\vert e^{-i\, {\hat {\sf H}}_k\, (-s)} \vert t\rangle
&=&\int\limits_{{\rm all}\; E}\! dE\; 
\vert \psi_{E}(t)\vert^2\, e^{iEs}\nn\\
&=&\l(\frac{1}{a^3(t)}\r)\!\! 
\int\limits_{{\rm all}\; E}\! dE\; \vert \chi_{E}(t)\vert^2\, 
e^{iEs},\qquad\label{eq:FKac}
\er
where the integral is all over the values of the energy~$E$.
Substituting this expression in the expression~(\ref{eq:PS1}) for 
the power spectrum, we obtain that
\beq
\left[k^3\; {\cal P}_{\Phi}(k)\right] 
=\l(\frac{k^{3}}{4 \pi^{2}\, a^3(t)}\r)
\int\limits_{-\infty}^{\infty}\! ds
\int\limits_{{\rm all}\; E}\! dE\;
\vert \chi_{E}(t)\vert^2\, e^{iEs}.\label{eq:psf}\nn\\
\eeq
On interchanging the order of the integrals and carrying the 
integral over $s$ first, we find that the power spectrum 
reduces to
\br
\left[k^3\; {\cal P}_{\Phi}(k)\right] 
&=&\l(\frac{k^{3}}{2 \pi \, a^3(t)}\r)\,
\int\limits_{{\rm all}\; E}\! dE\; \delta(E)\; 
\vert \chi_{E}(t)\vert^2
\label{eq:ps}\nn\\
&=& \l(\frac{k^{3}}{2 \, \pi \, a^3(t)}\r)\, 
\vert \chi_{_{0}}(t)\vert^2.
\er
Note that the integral over $s$ leads to a `density of states' that is
a delta function in $E$.  Evidently, this implies that only the wave
function corresponding to $E=0$ contributes to the sum, thereby
leading to the standard result one usually arrives at by the canonical
quantization procedure.  As we shall see later, path integral duality
introduces Planck scale modifications to the `density of states' , and
these modifications in turn lead to the quantum gravitational
corrections in the power spectrum.

We shall now use the above procedure to calculate the power spectrum
of the perturbations in de Sitter inflation.  Consider the case of
exponential inflation described by the scale factor
\beq
a(t)= e^{{\cal H}t},\label{eq:dsle}
\eeq
where ${\cal H}$ denotes the energy scale during inflation.  In such a
background, the Schrodinger equation~(\ref{eq:sch2}) reduces to
\beq
-\l(\frac{d^2 \chi_{E}}{dt^2}\r)
-\l[k^2\, e^{-2{\cal H}t}-\l(\frac{9\, {\cal H}^2}{4}\r)\r]\chi_{E}
=E\chi_{E}.\label{eq:sch-ds}
\eeq
The general solution to this differential equation is given by
\beq
\chi_{E}(t)= N_{E}\; H_{\nu}^{(1)}\!
\l(k{\cal H}^{-1}\, e^{-{\cal H} t}\r)
+M_{E}\; H_{\nu}^{(2)}\!
\l(k{\cal H}^{-1} e^{-{\cal H} t}\r),\label{eq:ds-sol}
\eeq
where $N_{E}$ and $M_{E}$ are $E$-dependent constants which are
determined by the initial conditions and the normalization
condition~(\ref{eq:norm2}).  The functions $H_{\nu}^{(1)}$ and
$H_{\nu}^{(2)}$ are the Hankel functions of the first and the second
kinds, respectively, and the quantity $\nu$ is given by
\beq
\nu=\l[\l(\frac{9}{4}\r)-\l(\frac{E}{{\cal H}^2}\r)\r]^{1/2}.
\label{eq:nu-ds}
\eeq

In inflationary cosmology, the initial conditions are imposed at very
early times when the modes are well within the Hubble radius.  Also,
it is assumed that, in the sub-Hubble limit (i.e. when $(k/a) \gg
{\cal H}$), the modes are in the vacuum state.  In the Schrodinger
picture that we are working with here, this condition corresponds to
choosing $\chi_{E}$ to be an `outgoing' wave at the `left infinity'.
For the case of de Sitter, this condition reduces to
\beq 
\lim_{t \to -\infty} \chi_{E}(t)
\propto \exp -\l(ik{\cal H}^{-1}\, e^{-{\cal H}t}\r)
\label{eq:icwf}
\eeq
which can be achieved by setting condition $M_{E}$ to zero in
Eq.~(\ref{eq:ds-sol}).  The quantity $N_{E}$ can be determined using
the normalization condition~(\ref{eq:norm2}).  We find that it is
given by (for details, see Appendix~\ref{sec:norma})
\beq
N_{E}
=\l\{\begin{array}{l}
\l[\frac{|{\rm sin}(\pi\nu)|}{4{\cal H}}\r]^{1/2},\;\;
\hfill{{\rm for}\;\; E < \l(9{\cal H}^2/4\r),} \\
\l[\frac{{\rm sinh}\l(\pi\vert\nu\vert\r)\, 
e^{-\l(2\pi\vert \nu\vert\r)}}
{4{\cal H}}\r]^{1/2},\;\;
\hfill{{\rm for}\;\; E \ge \l(9{\cal H}^2/4\r),} 
\label{eq:nc}
\end{array}\r.
\eeq
so that we have
\beq
\chi_{E}(t)
\!=\!\l\{\begin{array}{l}
\!\!\l[\frac{|{\rm sin}(\pi\nu)|}{4{\cal H}}\r]^{1/2} 
 H_{\nu}^{(1)}\!\l(k{\cal H}^{-1} e^{-{\cal H} t}\r),\\
\hfill{{\rm for}\,\, E < \l(9{\cal H}^2/4\r),}\\
\\
\!\!\l[\frac{{\rm sinh}\l(\pi\vert\nu\vert\r)\, 
e^{-\l(2\pi\vert \nu\vert\r)}} {4{\cal H}}\r]^{1/2}
H_{i\vert\nu\vert}^{(1)}\!\l(k{\cal H}^{-1} e^{-{\cal H} t}\r)\\
\hfill{{\rm for}\,\, E \ge \l(9{\cal H}^2/4\r).}\label{eq:ds-psol}
\end{array}\r.
\eeq
The wave function for $E=0$ is then given by
\beq
\chi_{_{0}}(t)
=\l(\frac{1}{\sqrt{4 {\cal H}}}\r)\, 
H_{(3/2)}^{(1)}\!\l(k{\cal H}^{-1} e^{-{\cal H}t}\r)
\label{eq:ds-E0}
\eeq
and the power spectrum (\ref{eq:ps}) can be written as
\beq
\left[k^3\; {\cal P}_{\Phi}(k)\right] 
=\l(\frac{k^{3}}{8\, \pi\, {\cal H}\, \, e^{-3{\cal H}t}}\r)\, 
\biggl\vert H_{(3/2)}^{(1)}\!
\l(k {\cal H}^{-1}\, e^{-{\cal H} t}\r)\biggr\vert^2.
\eeq
Using the expression for the Hankel function $H_{(3/2)}^{(1)}$ (see,
for instance, Refs.~\cite{texts}) it is then straightforward to obtain
the following spectrum at Hubble exit [i.e. when $\l(k{\cal H}^{-1}\,
e^{-{\cal H}t}\r)=1$]:
\beq
\left[k^3\; {\cal P}_{\Phi}(k)\right] 
=\l(\frac{{\cal H}^2}{2\, \pi^2}\r)\label{eq:psds}
\eeq
which is the standard, scale-invariant, spectrum one obtains in de
Sitter.  It should be pointed out here that, in obtaining the above
expression, we have evaluated the power spectrum when the modes leave
the Hubble radius.  In the literature, the spectrum of perturbations
is usually evaluated at the super-Hubble scales.  In the standard
theory, the power spectra at Hubble exit and at super-Hubble scales
typically differ in amplitude by a factor of order unity.

\section{Planck scale corrections due to path integral duality}
\label{sec:corrections}

In this section, using the approach of path integral duality, we shall
evaluate the Planck scale corrections to the scale invariant
perturbation spectrum~(\ref{eq:psds}).

To obtain the modified power spectrum, we shall use the path integral
duality modified Hadamard function~(\ref{eq:mHada}) in the
definition~(\ref{eq:psdfntn}) of the power spectrum.  Using the
expression~(\ref{eq:kernelfrw}) for the complete kernel, and the
Feynman-Kac formula~(\ref{eq:FKac}) for the kernel in the time
direction, it is straightforward to show that the modified spectrum is
given by
\br
\left[k^3\; {\cal P}_{\Phi}(k)\right]_{\rm (M)} 
\!\!\!&=&\!\!\l(\frac{k^{3}}{4 \pi^{2}\, a^3(t)}\r)
\int\limits_{-\infty}^{\infty}\! ds\; e^{i L_{_{\rm P}}^2/s}\nn\\
& &\qquad\qquad\quad\;\times\int\limits_{{\rm all}\; E}\! dE\;
\vert \chi_{E}(t)\vert^2\, e^{iEs}\nn\\
&=&\!\!\l(\frac{k^{3}}{4 \pi^2 \, a^3(t)}\r)
\int\limits_{{\rm all}\; E}\! dE\; \vert \chi_{E}(t)\vert^2\nn\\
& &\qquad\qquad\quad\;\times\int\limits_{-\infty}^{\infty}\!ds\; 
e^{i \l[(L_{_{\rm P}}^2/s) +E s\r]},\qquad\label{eq:psmod1}
\er
where, in the second expression, as we had done earlier, we have
interchanged the order of integration over~$E$ and~$s$.  On carrying
out the integral over $s$, we obtain the path integral duality
modified `density of states' to be (for details, see
Appendix~\ref{app:pidmds}):
\br
& &\!\!\!\!\!\!\!\!\int\limits_{-\infty}^{\infty}\!ds\; 
e^{i \l[(L_{_{\rm P}}^2/s) + E s\r]}\nn\\
& &\quad=(2\pi)\, \biggl[\delta(E) 
- \theta(E)\, \l(\frac{L_P}{\sqrt{E}}\r)\, 
J_{1}\!\l(2 L_{_{\rm P}}\sqrt{E}\r)\biggr],\qquad\label{eq:intsMod} 
\er
where $\theta(E)$ denotes the step function and $J_{1}$ is the Bessel
function of order unity.  On substituting this `density of states' in
the expression~(\ref{eq:psmod1}), we obtain the modified power
spectrum to be
\br
& &\!\!\!\!\!\!\!
\left[k^3\; {\cal P}_{\Phi}(k)\right]_{\rm (M)}\nn\\  
& &\;\;=\l(\frac{k^{3}}{2 \pi \, a^3(t)}\r)\, 
\vert \chi_{_0}(t)\vert^2\nn\\ 
& &\qquad-\l(\frac{L_{_{\rm P}} \, k^{3}}{2 \pi \, a^3(t)}\r)\, 
\int\limits_{0}^{\infty}\! \frac{dE}{\sqrt{E}}\, 
J_1\!\l(2 L_{_{\rm P}} \sqrt{E}\r)\vert \chi_{E}(t)\vert^2.\qquad
\label{eq:mpsge}
\er

The following points need to be emphasized regarding the expressions
we have obtained above for the modified `density of states' and the
corresponding power spectrum\footnote{We should stress here that the
quantity in Eq.~(\ref{eq:intsMod}) is not positive definite and,
therefore, it does not represent a genuine density of states.  We have
used the term `density of states' simply as a convenient shorthand for
referring to this quantity.}.  It is clear that it is the second term
in these expressions that leads to the Planck scale corrections in the
`density of states' and the primordial spectrum.  Importantly, as
required, this additional term vanishes in the limit of $L_{_{\rm P}}
\to 0$ so that we recover the standard result.

Before we proceed to evaluate the corrections, it is important that we
highlight another feature of the correction term in the above modified
spectrum, and make suitable clarifying remarks.  Note that, in the
expression~(\ref{eq:mpsge}) for the modified spectrum, it is wave
functions with energy $E>0$ that contribute to the corrections.  In
field theoretic language, wave functions with $E> 0$ correspond to
massive modes.  (Recall that the first term in~(\ref{eq:mpsge}) arises
due to the $E=0$, which corresponds to a massless mode, in
confirmation of the canonical picture.)  In other words, it is the
massive modes that lead to the Planck scale corrections and, in fact,
in our model, the corrections are a sum of the contributions due to
{\it all}\/ the massive modes.  But, it is well known that, in the
standard inflationary picture, the amplitude of massive modes decays
at super-Hubble scales.  Hence, within the conventional scenario, the
amplitude of the corrections would be expected to decay at the
super-Hubble scales.  However, we shall adopt the point of view that,
since, in our model, the massive modes represent the Planck scale
corrections to the standard, massless modes, the mechanism that
`freezes' the amplitude of the standard spectrum at super Hubble
scales will also `freeze' the amplitude of the Planck scale
corrections at their value at Hubble exit.  Therefore, in what
follows, we shall evaluate the corrections to the standard power
spectrum when the modes leave the Hubble radius.

Let us now evaluate the Planck scale corrections to the
spectrum~(\ref{eq:psds}) we had obtained in the last section for the
case of the de Sitter universe.  On substituting the wave
function~(\ref{eq:ds-psol}) in the expression~(\ref{eq:mpsge}) for the
modified power spectrum and imposing the condition of Hubble exit, we
obtain that
\beq
\left[k^3\; {\cal P}_{\Phi}(k)\right]_{\rm (M)} 
= \l(\frac{{\cal H}^2}{4\pi^2}\r) 
-{\cal C}(k,{\cal H},L_{_{\rm P}}),\label{ps-dsmod}
\eeq
where the term ${\cal C}(k,H,L_{_{\rm P}})$ denotes the Planck 
scale corrections to the scale invariant power spectrum and is 
given by
\br
& &\!\!\!\!\!\!\!\!\!
{\cal C}(k,{\cal H},L_{_{\rm P}})\nn\\ 
& &= \l(\frac{{\cal H}^2\, L_{_{\rm P}}\!}{8\pi}\r)\!\!
\int\limits_{0}^{(9{\cal H}^2/4)}\! \frac{dE}{\sqrt{E}}\; 
J_{1}\!\l(2 L_{_{\rm P}} \sqrt{E}\r)\, \sin(\pi \nu)\nn\\
& &\qquad\qquad\qquad\qquad\qquad\qquad\qquad\qquad
\times\;
\vert H_{\nu}^{(1)}\!\l(1\r)\vert^2\nn\\
& &\;\;
+\l(\frac{{\cal H}^2\, L_{_{\rm P}}\!}{8\pi}\r)\!\!
\int\limits_{(9{\cal H}^2/4)}^{\infty}\! \frac{dE}{\sqrt{E}}\; 
J_{1}\l(2 L_{_{\rm P}} \sqrt{E}\r)\, \sinh(\pi \vert\nu\vert)\nn\\
& &\qquad\qquad\qquad\qquad\qquad\;\times\;
e^{-\l(2 \pi\, \vert\nu\vert\r)}\;
\vert H_{\nu}^{(1)}\!\l(1\r)\vert^2.\quad
\er
Clearly, the corrections are independent of $k$.  Therefore, the
corrections do not result in any features in the standard, scale
invariant power spectrum, but they simply change the amplitude of the
spectrum\footnote{In Appendix \ref{app:limH0}, as a consistency check,
we have explicitly shown that, in the limit of $H \to 0$, the modified
spectrum evaluated at a given time reduces to the expected power
spectrum of a scalar field in flat space-time.}.  In fact, in a de
Sitter universe, this feature seems to be common to many of the high
energy models that have been considered in the literature---for
example, see, Refs.~\cite{general,Hofmann:2004}.  Since the amplitude
cannot be determined from the observations, the above corrections are,
in principle, unobservable.  Nevertheless, the amplitude of the Planck
scale corrections is of theoretical interest and, in what follows, we
shall evaluate this amplitude.

In terms of the variable
\beq
{\cal E}=\l(2\, L_{_{\rm P}}\, \sqrt{E}\r),
\eeq
we find that the corrections ${\cal C}(k,{\cal H},L_{_{\rm P}})$ 
can be written as
\beq
{\cal C}(k,{\cal H},L_{_{\rm P}})
=\l(\frac{{\cal H}^2}{8\pi}\r)\, 
\l[{\cal A}(\delta)+{\cal B}(\delta)\r],
\eeq
where the quantities ${\cal A}$ and ${\cal B}$ depend only on 
the dimensionless quantity $\delta$ defined to be
\beq
\delta=\l({\cal H}\, L_{_{\rm P}}\r)=\l({\cal H}/M_{_{\rm P}}\r)
\eeq
with $M_{_{\rm P}}$ being the Planck mass.
The quantities ${\cal A}$ and ${\cal B}$ are described by the 
integrals
\br
{\cal A}(\delta)
\!&=&\!\!\!\int\limits_{0}^{(3\, \delta)}\! d{\cal E}\, 
J_{1}\!\l({\cal E}\r)\, \sin(\pi \nu)\;
\vert H_{\nu}^{(1)}\!\l(1\r)\vert^2,\\
{\cal B}(\delta)
\!&=&\!\!\!\int\limits_{(3\,\delta)}^{\infty}\! d{\cal E}\, 
J_{1}\!\l({\cal E}\r)\, \sinh(\pi \vert\nu\vert)\;
e^{-\l(2 \pi\, \vert\nu\vert\r)}\,
\vert H_{\nu}^{(1)}\!\l(1\r)\vert^2\qquad
\er
and the quantity $\nu$ can be expressed in terms of ${\cal E}$ and 
$\delta$ as follows:
\beq
\nu
=\l[\l(\frac{9}{4}\r)
-\l(\!\frac{\cal E}{2\,\delta}\!\r)^2\r]^{1/2}.
\eeq
However, closed form expressions for the above integrals describing
${\cal A}$ and ${\cal B}$ do not seem to exist.  Therefore, we have
evaluated them numerically and, in the figure below, we have plotted
the sum of ${\cal A}$ and ${\cal B}$ as a function of $\delta$.
\begin{figure}[!htb]
\begin{center}
\epsfxsize 3.30 in
\epsfysize 2.70 in
\epsfbox{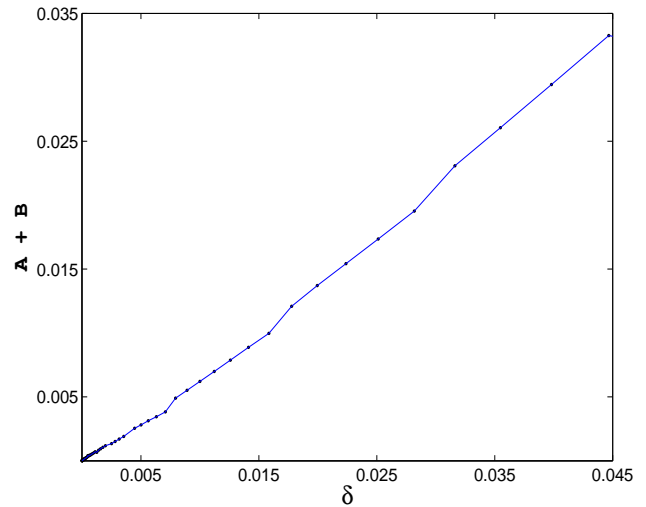}
\caption{$\l({\cal A}+{\cal B}\r)$ plotted as a function of the 
quantity $\delta$.}
\end{center}
\end{figure}
It is evident from the above figure that, in a de Sitter universe, 
the Planck scale corrections to primordial spectrum are linear 
in $\delta$.

\section{Summary and discussion}\label{sec:discussion}

In this work, using the locally Lorentz invariant approach of path
integral duality, we have evaluated the Planck scale corrections to
the primordial perturbation spectrum in the case of exponential
inflation.  We find that the Planck scale corrections are of the order
of $\l({\cal H}/M_{_{\rm P}}\r)$.  Moreover, the corrections are
completely scale-free, which implies that they simply change the
amplitude of the power spectrum.  As we pointed out in the last
section, the amplitude in itself cannot be determined from the
observations and, as a result, the corrections we have obtained are
unobservable.  However, the amplitude of the Planck scale corrections
(whether they would depend on, say, $({\cal H}/M_{_{\rm P}})$ or
$({\cal H}/M_{_{\rm P}})^{2}$) have been of theoretical interest and,
our calculation assumes importance in this context.

At this stage, it is interesting to compare our results with that of
Ref.~\cite{chouha05}.  As we had mentioned in the introduction, in
this work, the authors evaluate the power spectrum of the
gravitational waves in exponential inflation.  For computational
convenience, they use a particular modified dispersion relation to
mimic the high energy effects.  For the case of the minimal energy
initial state---which is similar to the vacuum state we have chosen in
Eq.~(\ref{eq:icwf})---they obtain a spectrum which has a blue tilt
with superimposed oscillations.  In contrast, using the locally
Lorentz invariant proper time method, we obtain a spectrum that is
strictly scale invariant with the Planck scale corrections of order
$\l({\cal H}/M_{_{\rm P}}\r)$.

In cosmological perturbation theory, it is well-known that, the metric
perturbations do not couple to the matter perturbations in a de Sitter
background (see, for instance, Ref.~\cite{mukhanov92}).  For this
reason, the exact de Sitter case is considered pathological and,
ideally, one would like to carry out the analysis in slow-roll
inflation, or, at least for the case of power-law inflationary
scenario.  However, the wave functions $\chi_{E}$ for $E\ne 0$ are not
known in an analytic form in such cases and one needs to resort to a
WKB approximation to evaluate the wave functions and the resulting
corrections.  We hope to address this issue in a forthcoming
publication.

\section*{Acknowledgments}

The authors would wish to thank T.~Padmanabhan for discussions and
comments on the manuscript. LS would like to thank the High Energy,
Cosmology and Astroparticle Physics Group of The Abdus Salam
International Centre for Theoretical Physics, Trieste, Italy for
hospitality, where part of this work was carried out.

\appendix


\section{Proper time representation of the Hadamard function}
\label{app:hadamard}

In this appendix, we shall outline as how the Hadamard function in
flat space-time can be written as in Eq.~(\ref{eq:sub1}).
Let us begin by recalling that the Hadamard function of a quantized, 
scalar field ${\hat \Phi}$ is defined as follows:
\beq
G^{(1)}({\tilde x},{\tilde x}')
= \langle 0\vert\, {\hat \Phi}({\tilde x})\; {\hat \Phi}({\tilde x'}) 
+{\hat \Phi}({\tilde x'})\; {\hat \Phi}({\tilde x})\, \vert 0\rangle.
\eeq
Consider a massive, scalar field in flat space-time.
Using the canonical quantization procedure, it is straightforward to 
show that the Hadamard function is given by
\br
G^{(1)}({\tilde x},{\tilde x}') 
\!\!&=&\!\!\int\! \frac{d^3{\bf k}}{(2\pi)^3}\, 
\l(\!\frac{1}{2\omega}\!\r)  
\l(e^{-i\omega(t-t')}+e^{i\omega(t-t')}\r)\nn\\
& &\qquad\qquad\qquad\qquad\qquad\times\;\,
e^{i{\bf k}\cdot({\bf x} - {\bf x}')},\qquad
\er
where $\omega^2=\l(k^2+m^2\r)$.
This expression can, in turn, be written as
\br
G^{(1)}({\tilde x},{\tilde x}') 
\!\!&=&\!\! \int\! \frac{d^3{\bf k}}{(2\pi)^3}\, 
\int\!\frac{dk_{0}}{(2k_{0})}\,  
\l[\delta(k_{0}-\omega) +\delta(k_{0}+\omega)\r]\nn\\
& &\qquad\qquad\qquad\qquad\;\times\;
e^{-i\l[k_{0}(t-t')-{\bf k}\cdot({\bf x} - {\bf x}')\r]}\nn\\
&=&\!\!\int\! \frac{d^4{\tilde k}}{(2\pi)^3}\; 
\delta\l({\tilde k}^2-m^2\r)\;  
e^{-i{\tilde k}\cdot({\tilde x} - {\tilde x}')},\label{eq:Had-FT-def}
\er
where ${\tilde k}\equiv k^{\mu}$, ${\tilde k}^2\equiv \l(k_{\mu}
k^{\mu}\r)$, $\l({\tilde k}\cdot{\tilde x}\r)\equiv \l(k_{\mu}\, 
x^{\mu}\r)$, and in the final expression we have made use of the 
relation 
\br
\l(\!\frac{1}{2\omega}\!\r)\, \l[\delta\l(k_{0}-\omega\r)
+\delta\l(k_{0}+\omega\r)\r]
&=&\delta\l(k_{0}^2-\omega^2\r)\nn\\
&=&\delta\l({\tilde k}^2-m^2\r).\qquad
\er
On using the integral representation of the $\delta$ function, we 
obtain that 
\br
G^{(1)}({\tilde x},{\tilde x}') 
&=& \int\limits_{-\infty}^{\infty}\!ds\, e^{-i m^{2} s}
\int\! \frac{d^4{\tilde k}}{(2\pi)^4}\, 
e^{i{\tilde k}^2 s}\, 
e^{-i{\tilde k}\cdot({\tilde x} - {\tilde x}')}\nn\\
&=& \int\limits_{-\infty}^{\infty}\! ds\, e^{-im^{2} s}\,
K({\tilde x},{\tilde x}';s),\label{eq:Had-Intrep-def}
\er
where in the last expression we have identified the integral over 
${\tilde k}$ to be the path-integral kernel of a free particle in
four dimensions.


\section{Determining the normalization 
constant~$N_{E}$}\label{sec:norma}

In this appendix, we shall evaluate the normalization constant 
$N_{E}$ for the wave functions corresponding to the Hamiltonian 
${\sf H}_{k}$ in de Sitter. 
Recall that, in a de Sitter background, for the initial conditions 
of our interest, we can set $M_{E}$ to zero in Eq.~(\ref{eq:ds-sol}), 
so that the wave function $\chi_{E}$ is given by 
\beq
\chi_{E}(t)
= N_{E}\; H_{\nu}^{(1)}\!\l(k{\cal H}^{-1}\, e^{-{\cal H} t}\r),
\label{eq:ds-sol1}
\eeq
On substituting this wave function in the left-hand side of 
Eq.~(\ref{eq:norm2}), we get 
\br
\!\!\!\!
{\cal I}(E,E') 
\!&=&\! \l(N_{E}\, N_{E'}^{*}\r)\, 
\int\limits_{-\infty}^{\infty}\! dt\;
H_{\nu}^{(1)}\!\l(k{\cal H}^{-1}\, e^{-{\cal H} t}\r)\nn\\
& &\qquad\qquad\qquad\qquad\times\;
H_{\nu'}^{(1)*}\!\!\l(k{\cal H}^{-1}\, e^{-{\cal H} t}\r)\nn\\
&=&\! \l(N_{E}\, N_{E'}^{*}\r)\, 
\int\limits_{0}^{\infty}\!\frac{d \eta}{{\cal H} \eta}\;  
H_{\nu}^{(1)}(k\eta)\; H_{\nu'}^{(1)*}(k \eta),\quad
\label{eq:lhs1}
\er
where, in the second expression, we have expressed the integral
in terms of the new independent variable $\eta = \l({\cal H}\, 
{\rm e}^{{\cal H}t}\r)^{-1}$.
Also, the quantity $\nu'$ denotes the quantity $\nu$ [defined 
in Eq.~(\ref{eq:nu-ds})] corresponding to the energy $E'$.

\begin{widetext}
In order to establish the normalization condition~(\ref{eq:norm2}) 
and arrive at the value of normalization constant $N_{E}$, we shall 
adopt the following procedure. 
We shall relate the Hankel functions appearing in the
integral~(\ref{eq:lhs1}) above to the MacDonald function 
$K_{\nu}$ and use the following standard integral (see, 
for example, Ref.~\cite{gandr80}, p.~693):
\br
\int\limits_{0}^{\infty}\! du\, u^{-\lambda}\, 
K_{\alpha}(p u)\, K_{\beta}(q u) 
\!\!&=&\!\! \l(\frac{2^{-(2+\lambda)}}{\Gamma(1 - \lambda)}\r)\, 
p^{-(1+\beta-\lambda)}\, q^{\beta}\nn\\
& &\times\;
\Gamma\l(\frac{1-\lambda +\alpha +\beta}{2}\r)\; 
\Gamma\l(\frac{1-\lambda -\alpha-\beta}{2}\r)\;
\Gamma\l(\frac{1-\lambda -\alpha +\beta}{2}\r)\; 
\Gamma\l(\frac{1-\lambda +\alpha -\beta}{2}\r)\nn\\
& &\times\; 
F\l[\l(\frac{1 -\lambda +\alpha +\beta}{2}\r),
\l(\frac{1 -\lambda -\alpha +\beta}{2}\r), \l(1 - \lambda\r), 
\l(\frac{a^2-b^2}{a^{2}}\r)\r],\label{eq:asp676}
\er
where $\Gamma$ denotes the Gamma function and $F$ is the 
hypergeometric function.
\end{widetext}

Note that, while the argument of the Hankel function appearing 
in the integral~(\ref{eq:lhs1}) is always real, the index $\nu$ 
is real for $E<(9H^2/4)$, but it is imaginary for $E>(9H^2/4)$. 
For real values of the index, the Hankel functions $H_{\nu}^{(1)}$ 
and $H_{\nu}^{(2)}$ are complex conjugates of each other, and 
they are related to the MacDonald functions $K_{\nu}$ as follows 
(cf.~Ref.~\cite{abramowitz64}, p.~375)
\begin{subequations}
\label{eq:HKrelationsrnu}
\br
H_{\nu}^{(1)}(z)
&=& -\l(\frac{2 i}{\pi}\r)\, e^{-(i\pi\nu/2)}\; K_{\nu}(-iz),\\
H_{\nu}^{(2)}(z)
&=& \l(\frac{2 i}{\pi}\r)\, e^{(i\pi\nu/2)}\; K_{\nu}(iz).
\er
\end{subequations}
Whereas, when $\nu$ is imaginary, the corresponding relations 
can be obtained to be
\begin{subequations}
\br
H_{i\vert\nu\vert}^{(1)}(z)
&=& -\l(\frac{2 i}{\pi}\r)\, e^{(\pi\vert\nu\vert/2)}\, 
K_{i\vert\nu\vert}(-iz),\label{eq:H1Kinu}\\
H_{i\vert\nu\vert}^{(2)}(z)
&=& \l(\frac{2 i}{\pi}\r)\, e^{-(\pi\vert\nu\vert/2)}\, 
K_{i\vert\nu\vert}(iz).\label{eq:H2Kinu}
\er
\end{subequations}
For an imaginary index, the two Hankel functions are {\it not}\/
complex conjugates of each other, but are related as follows:
\beq
H_{i\vert\nu\vert}^{(1)*}(z)
=e^{(\pi\nu)}\, H_{i\vert\nu\vert}^{(2)}(z).
\label{eq:H1H2inu}
\eeq

Let us first consider the case wherein $E$ and $E'$ are both 
greater than~$(9H^2/4)$ so that $\nu$ and $\nu'$ are both 
imaginary.
In such a case, on using the relations~(\ref{eq:H1Kinu}) 
and~(\ref{eq:H1H2inu}), the integral~(\ref{eq:lhs1}) can 
be expressed as
\br
{\cal I}(E,E')
\!\!&=&\!\!
\l(\frac{4\, N_{E}\, N_{E'}^{*}}{\pi^2\, {\cal H}}\r)\, 
e^{\l[\pi(\vert\nu\vert+\vert\nu'\vert)/2\r]}\nn\\
& &\qquad\times\;\int\limits_{0}^{\infty}\! \frac{d \eta}{\eta}\;  
K_{i\vert\nu\vert}(-ik\eta)\; K_{i\vert\nu'\vert}(ik\eta).\qquad
\er
In order to identify the divergent part in the limit of $E\to E'$, 
we shall express this integral as a limit in the following fashion:
\br
& &\!\!\!\!\!\!\!\!\!
{\cal I}(E,E')\nn\\
& &= \l(\frac{4\, N_{E}\, N_{E'}^{*}}{\pi^2\, {\cal H}}\r)\, 
e^{\l[\pi(\vert\nu\vert+\vert\nu'\vert)/2\r]}\nn\\
& &\quad\times \lim_{\epsilon\to 0}\;
\int\limits_{0}^{\infty}\! d \eta\, \eta^{-(1-2\epsilon)}\; 
K_{i\vert\nu\vert}(-i k \eta)\; 
K_{i\vert\nu'\vert}(i k \eta).\qquad\label{eq:lhs3}
\er
On comparing Eqs.~(\ref{eq:lhs3}) and~(\ref{eq:asp676}), we 
can identify that
\br
\lambda &=& \l(1 - 2 \epsilon\r),\nn\\
p &=& - (i k),\qquad q = (i k),\nn\\
\alpha &=& \l(i\vert\nu\vert\r),\qquad
\beta = \l(i\vert\nu'\vert\r),
\er
so that we have 
\begin{widetext}
\br
{\cal I}(E,E')
&=&\l(\frac{N_{E}\, N_{E'}^{*}}{2\,\pi^2\, {\cal H}}\r)\, 
e^{\l[\pi(\vert\nu\vert+\vert\nu'\vert)/2\r]}\; e^{\l(\pi\nu'\r)}\;
\Gamma\l[\frac{i(\vert\nu\vert + \vert\nu'\vert)}{2}\r]\;
\Gamma\l[\frac{-i(\vert\nu\vert + \vert\nu'\vert)}{2}\r]\nn\\ 
& &\qquad\qquad\qquad\qquad\qquad\qquad\qquad\qquad
\times\; \lim_{\epsilon \to 0}\;
\l[\frac{\Gamma\l(\epsilon 
+ \l[i\,(\vert\nu\vert -\vert\nu'\vert)/2\r]\r)\;
\Gamma\l(\epsilon -\l[i\,(\vert\nu\vert 
- \vert\nu'\vert)/2\r]\r)}{\Gamma(2 \epsilon)}\r],\qquad\qquad
\label{eq:lhs4}
\er
\end{widetext}
where we have made use of the fact that $F(a,b,c,0)=1$ (cf. 
Ref.~\cite{abramowitz64}, p.~556).
On using the following, standard properties of $\Gamma$ function 
(cf. Ref.~\cite{abramowitz64}, p.~256)
\begin{subequations}
\label{eq:Gammafnp}
\br
\Gamma(1+z) &=& z\, \Gamma(z),\\
\Gamma(z)\, \Gamma(1-z) 
&=& -z\, \Gamma(z)\, \Gamma(-z)\nn\\
&=& \l(\frac{\pi}{{\rm sin}\l(\pi z\r)}\r),\\
\vert \Gamma(iz)\vert^2
&=& \l(\frac{\pi}{z\; {\rm sinh}\l(\pi z\r)}\r),\\
\vert\Gamma(1+iz)\vert^2
&=&\l(\frac{\pi\, z}{{\rm sinh}\l(\pi z\r)}\r),
\qquad\qquad\er
\end{subequations}
we find that the expression~(\ref{eq:lhs4}) can be expressed as
\br
& &\!\!\!\!\!\!
{\cal I}(E,E')\nn\\
& &=\l(\frac{N_{E}\, N_{E'}^{*}}{\pi\, {\cal H}}\r)\! 
\l(\frac{e^{\l[\pi(\vert\nu\vert
+3\, \vert\nu'\vert)/2\r]}}{\l[(\vert\nu\vert
+\vert\nu'\vert)/2\r]\; {\rm sinh}\l[\pi
(\vert\nu\vert+\vert\nu'\vert)/2\r]}\r)\nn\\ 
& &\qquad\qquad\qquad\;
\times\; \l(\frac{\l[\pi\, \l(\vert\nu\vert
-\vert\nu'\vert\r)/2\r]}{{\rm sinh}\l[\pi
\l(\vert\nu\vert-\vert\nu'\vert\r)/2\r]}\r)\nn\\
& &\qquad\qquad\qquad\;
\times\,\lim_{\epsilon \to 0}\;
\l(\frac{\epsilon}{\epsilon^2 
+ \l[\l(\vert\nu\vert - \vert\nu'\vert\r)/2\r]^{2}}\r).\;\;
\label{eq:lhs5}
\er
Using the following representation of the delta function
\be
\delta(x) 
= \l(\frac{1}{\pi}\r)\; \lim_{\epsilon \to 0}\; 
\l(\frac{\epsilon}{\epsilon^2 + x^2}\r),\label{eq:delrep}
\eeq
we can then write that 
\beq
{\cal I}(E,E')
=\l(\frac{2\, \vert N_{E}\vert^{2}}{{\cal H}}\r)\,
\l(\frac{e^{\l(2\pi \vert\nu\vert\r)}}{\vert\nu\vert\,
{\rm sinh}\l(\pi\vert\nu\vert\r)}\r)\;
\delta(\vert\nu\vert-\vert\nu'\vert).\label{eq:lhs6}
\eeq
From the relation between $E$ and $\nu$ [cf. Eq.~(\ref{eq:nu-ds})],
we find that
\br
\delta(E-E') 
&=&\l(\frac{1}{{\cal H}^2}\r) \delta\l(\nu^2 - \nu'^2\r)\nn\\
&=& \l(\frac{1}{2\, {\cal H}^2\, \vert \nu\vert}\r) 
\delta\l(\vert\nu\vert - \vert\nu'\vert\r).\label{eq:rhs1}
\er
Therefore, we have
\beq
{\cal I}(E,E')
=\l(4\, {\cal H}\, \vert N_{E}\vert^{2}\r)\,
\l(\frac{e^{\l(2\pi 
\vert\nu\vert\r)}}{{\rm sinh}\l(\pi\vert\nu\vert\r)}\r)\;
\delta(E-E')
\eeq
which, finally, leads to
\beq
\vert N_{E}\vert^{2}
=\l(\frac{1}{4\, {\cal H}}\r)\, 
{\rm sinh}\l(\pi\vert\nu\vert\r)\; 
e^{-\l(2\pi \vert\nu\vert\r)}.\label{eq:NEinu}
\eeq

Let us now assume that both $E$ and $E'$ are less that $(9H^2/4)$ 
so that $\nu$ as well as $\nu'$ are real quantities.
As we had done earlier, we shall express the integral~(\ref{eq:lhs1}) 
as a limit as follows:
\br
\!\!\!\!\!\!\!\!\!
{\cal I}(E,E')
&=&\l(\frac{4\, N_{E}\,N_{E'}^{*}}{\pi^2\, {\cal H}}\r)\, 
e^{-\l[i\pi(\nu-\nu')/2\r]}\nn\\
& &\times \int\limits_{0}^{\infty}\! d \eta\,
 \eta^{-(1-2i\epsilon)}\,
K_{\nu}(-i k \eta)\, K_{\nu'}(i k \eta).\quad\label{eq:lhs7}
\er
where we have made use of the relations~(\ref{eq:HKrelationsrnu}).
On comparing Eqs.~(\ref{eq:lhs7}) and~(\ref{eq:asp676}), we 
can identify that
\beq
\lambda = \l(1 - 2 i \epsilon\r),\;\,
a = - (i k),\;\; b = (i k),\;\,
\alpha = \nu\;\,{\rm and}\;\, \beta = \nu',\label{eq:comp}
\eeq
so that we can write 
\begin{widetext}
\br
{\cal I}(E,E')
&=&\l(\frac{N_{E}\, N_{E'}^{*}}{2\, \pi^2\, {\cal H}}\r)\, 
e^{-\l[i\pi(\nu-3\, \nu')/2\r]}\;
\Gamma\l[\frac{\nu + \nu'}{2}\r]\;
\Gamma\l[-\l(\frac{\nu + \nu'}{2}\r)\r]\nn\\ 
& &\qquad\qquad\qquad\qquad\qquad\qquad\qquad\times\;
\lim_{\epsilon \to 0}\; 
\l[\frac{\Gamma\l(i \epsilon + [(\nu - \nu')/2]\r)\, 
\Gamma\l(i\epsilon - [(\nu - \nu')/2]\r)}{\Gamma(2 i \epsilon)}\r], 
\er
where, again, we have made use of the result that $F(a,b,c,0)=1$. 
On using the properties~(\ref{eq:Gammafnp}) of 
the $\Gamma$ function, we find that the above expression can be 
written as
\br
{\cal I}(E,E')
&=&\l(\frac{N_{E}\, N_{E'}^{*}}{\pi\, {\cal H}}\r)\! 
\l(\frac{i\, e^{-\l[i\pi(\nu+3\, \nu')/2\r]}}{\l[(\nu+\nu')/2\r]\; 
{\rm sin}\l[\pi(\nu+\nu')/2\r]}\r)\nn\\ 
& &\qquad\qquad\qquad\quad
\times\;\Gamma\l(1+\l[\l(\nu-\nu'\r)/2\r]\r)\;
\Gamma\l(1-\l[\l(\nu-\nu'\r)/2\r]\r)
\times\,\lim_{\epsilon \to 0}\;
\l(\frac{\epsilon}{\epsilon^2 
+ \l[\l(\nu -\nu'\r)/2\r]^{2}}\r)\nn\\
&=&\l(\frac{2\, \vert N_{E}\vert^{2}}{{\cal H}}\r)\,
\l(\frac{i\, e^{-\l(2i\pi \nu\r)}}{\nu\;
{\rm sin}\l(\pi\, \nu\r)}\r)\;
\delta(\nu-\nu')
=\l(4\,{\cal H}\, \vert N_{E}\vert^{2}\r)\,
\l(\frac{i\, e^{-\l(2i\pi \nu\r)}}{\nu\;
{\rm sin}\l(\pi\, \nu\r)}\r)\; \delta(E-E'),
\er
\end{widetext}
where in arriving at the second and the third equalities, we have 
made use of the relations~(\ref{eq:delrep}) and~(\ref{eq:rhs1}), 
respectively.
Dropping irrelevant phase factors, we obtain
\beq
\vert N_{E}\vert^2 
= \l(\frac{1}{4 {\cal H}}\r)\, \vert\sin(\pi \nu)\vert.
\label{eq:NErnu}
\eeq

We have quoted the results~(\ref{eq:NEinu}) and~(\ref{eq:NErnu}) 
in Eq.~(\ref{eq:nc}).


\section{Path integral duality modified `density of states'}
\label{app:pidmds}

In this appendix, we outline the derivation of the result we have
quoted in Eq.~(\ref{eq:intsMod}). 
Let
\beq
{\cal G}\l(E, L_{_{\rm P}}\r)
=\int\limits_{-\infty}^{\infty}\!ds\; 
e^{i \l[(L_{_{\rm P}}^2/s) + E s\r]}
\eeq
which can be expressed as
\br
{\cal G}\l(E, L_{_{\rm P}}\r)
&=&\int\limits_{0}^{\infty}\!ds\; 
e^{i \l[(L_{_{\rm P}}^2/s) + E s\r]}
+ \int\limits_{0}^{\infty}\!ds\; 
e^{-i \l[(L_{_{\rm P}}^2/s) + E s\r]}\nn\\
&=& \lim_{\epsilon \to 0^{+}} 
\int\limits_{0}^{\infty}\! ds \,
e^{-\l[(\epsilon-iE)s -(i\, L_{_{\rm P}}^2/s)\r]}\nn\\
& &\quad\;\;
+\lim_{\epsilon \to 0^{+}} 
\int\limits_{0}^{\infty}\! ds \,
e^{-\l[(\epsilon+iE)s + (i L_{_{\rm P}}^2/s)\r]},
\er
where we have introduced a small parameter $\epsilon$ which we 
shall eventually set to zero. 
On using the following integral representation of the modified 
Bessel function (cf. Ref.~\cite{gandr80}, p.~376)
\beq
\int\limits_{0}^{\infty}\! du\; \exp{-\l[a\, u + (b/u)\r]}
=\sqrt{\frac{4\, b}{a}}\;\, K_{1}\!\l(\sqrt{4ab}\r)
\label{eq:intrepK}
\eeq
which is valid for ${\rm Re.}~a > 0$ and ${\rm Re.}~(b) \ge 0$,
we obtain that

\br
& &\!\!\!\!\!\!\!\!\!\!\!
{\cal G}\l(E, L_{_{\rm P}}\r)\nn\\
&=& \lim_{\epsilon \to 0^{+}} \, 
2 \l(\frac{-i \, L_{_{\rm P}}^2}{\epsilon -iE}\r)^{1/2}
K_{1}\l(2\l[-i L_{_{\rm P}}^2 \l(\epsilon -iE\r)\r]^{1/2}\r)\nn\\
&+&\lim_{\epsilon \to 0^{+}}\, 
2 \l(\frac{i \, L_{_{\rm P}}^2}{\epsilon +iE}\r)^{1/2}
K_{1}\l(2\l[i L_{_{\rm P}}^2 \l(\epsilon +iE\r)\r] ^{1/2}\r),\qquad
\label{eq:d3}
\er
We shall now utilize the following series representation of the 
Macdonald function~$K_{1}$ (cf. Ref.~\cite{abramowitz64}, p.~375):
\br
\!\!\!\!\!\!\!\!
K_{1}(z) 
\!\!&=&\!\! \l(\frac{1}{z}\r) + \ln\l(\frac{z}{2}\r)\, I_{1}(z)\nn\\ 
& &\!\!\!\!
-\l(\frac{z}{4}\r)\,
\sum\limits_{k = 0}^{\infty} 
\l(\frac{\psi(k + 1) + \psi(k + 2)}{k! (k + 1)!}\r)\!
\l(\frac{z}{2}\r)^{2k}\!\!\!,\quad\label{eq:serrepK}
\er
where $I_1$ is the modified Bessel function of order unity and $\psi$
denotes the di-gamma function.  On using this series representation
and the relation $I_{1}(z)=-\l[i\, J_{1}(i z)\r]$ ($J_{1}$ being the 
Bessel function), we find that we can write the modified `density of 
states' as follows:
\br
\!\!
{\cal G}\l(E, L_{_{\rm P}}\r)
\!\!&=&\!\!\! \lim_{\epsilon \to 0^{+}}
\l(\frac{2\, \epsilon}{\epsilon^{2}+E^{2}}\r)
+ \l(\frac{2\pi i L_{_{\rm P}}}{\sqrt{E}}\r)
I_{1}\!\l(2 i L_{_{\rm P}}\sqrt{E}\r)\nn\\ 
&=&\!\! (2 \pi)\, \l[\delta(E)  - \l(\frac{L_P}{\sqrt{E}}\r)\,
J_1\!\l(2 L_{_{\rm P}}\, \sqrt{E}\r)\r],\;\;
\er
where we have also made use of the representation~(\ref{eq:delrep}) of
the delta function. This is the result~(\ref{eq:intsMod}) we have made
use in the text.


\section{A consistency check---Power spectrum in the limit of 
${\cal H} \to 0$}\label{app:limH0}

As a consistency check, in this appendix, we shall explicitly show
that, in the limit of $H \to 0$, the modified power spectrum  in
de Sitter reduces to that of the corresponding modified spectrum 
for a scalar field in flat space-time. 
In order to do so, let us first evaluate the modified spectrum in 
flat space-time.

In flat space-time (i.e. when $a = 1$), the normalized wave 
functions corresponding to the Schrodinger equation~(\ref{eq:sch1}) 
are given by
\beq
\chi_{E}(t)=\l(\frac{1}{4\pi \sqrt{k^{2}+E}}\r)^{1/2}\, 
e^{-i\, \sqrt{k^{2}+E}\; t}.\label{eq:fswf}
\eeq
On subsituting this wave function in the 
expression~(\ref{eq:mpsge}) for the modified power spectrum, 
we obtain that
\br
& &\!\!\!\!\!
\left[k^3\; {\cal P}_{\Phi}(k)\right]_{\rm (M)}\nn\\ 
& &\;=\l(\frac{k^{2}}{8 \pi^2}\r)\, 
- \l(\frac{L_{_{\rm P}} \, k^{3}}{8 \pi^{2}}\r)\, 
\int\limits_{0}^{\infty}\! \frac{dE}{\sqrt{E\, (k^{2}+E)}} \, 
J_1\!\l(2 L_{_{\rm P}}\, \sqrt{E}\r)\nn\\
& &\;=\l(\frac{k^{2}}{8 \pi^2}\r)\, 
\l[1- 2\, e^{-k L_{_{\rm P}}} \,
{\rm sinh}\l(k L_{_{\rm P}}\r)\r]\nn\\ 
& &\;=\l(\frac{k^{2}}{8 \pi^2}\r)\, e^{-2k L_{_{\rm P}}}
\label{eq:FSTps}
\er
This expression illustrates the fact that the duality principle
leads to an exponential suppression of power for the 
trans-Planckian modes (i.e. modes for which $k \gg L_{_{\rm P}}^{-1}$).

We shall now outline as to how the modified power spectrum in 
exponential inflation reduces to the above flat space-time 
expression in the limit of ${\cal H} \to 0$. 
In this limit, we find that the wave function~(\ref{eq:ds-psol}) 
in de Sitter space reduces to
\br
\!\!\!\!\!\!\!\!\!\!\!\!
\chi_{E}(t)
&\simeq& \l(4\, {\cal H}\r)^{-1/2}\,
e^{-\l(\pi\sqrt{E}/2\, {\cal H}\r)}\nn\\ 
& &\qquad\qquad\quad\times\;
H_{\l(i\sqrt{E}/{\cal H}\r)}^{(1)}\!\l(k {\cal H}^{-1}\, 
e^{-{\cal H}t}\r).\label{eq:dswffsl}
\er
Also, these wave functions are valid in the entire energy range of
$0 < E < \infty$.
For large complex order and real argument, the asymptotic 
form of the Hankel function $H_{i\nu}^{(1)}(z)$ is given by 
(cf.~Ref.~\cite{watson}, p.~263)
\beq
H_{i\nu}^{(1)}(z) 
\simeq \l[-\l(\pi \nu i/2\r)\, {\rm tanh}\, \gamma\r]^{-1/2}\;
e^{\nu\l({\rm tanh}\, \gamma -\gamma\r)}\, e^{-i\pi /4}.
\eeq
where $\nu=\l(z\; {\rm cosh}\, \gamma\r)$, $\gamma=\l(\alpha
+i\beta\r)$ and $0< \beta < \pi$.
If we choose $\beta=(\pi/2)$, then, we find that $\nu=\l(iz\; 
{\rm sinh}\, \alpha\r)$ and ${\rm tanh}\, \gamma={\rm coth}\, 
\alpha$.
In our case, we have $\nu=(i\sqrt{E}/H)$, $z=(k\, {\cal H}^{-1}\, 
e^{-{\cal H}t})$ so that ${\rm sinh}\, \alpha
=(\sqrt{E}\, k^{-1}\, e^{-{\cal H}t})$.
On using these expressions in the above asymptotic expansion of 
the Hankel function, we find that, as ${\cal H}\to 0$,
\begin{widetext}
\beq
\left \vert 
H_{\l(i\sqrt{E}/{\cal H}\r)}^{(1)}\!\l(k {\cal H}^{-1}\, 
e^{-{\cal H}t}\r)\right\vert^{2}
\to \l(\frac{{\cal H}\, {\rm tanh}\, \alpha}{\pi \sqrt{E}}\r)\, 
e^{\l(\pi\sqrt{E}/{\cal H}\r)}
=\l(\frac{{\cal H}}{\pi \sqrt{k^{2}+E}}\r)\, 
e^{\l(\pi\sqrt{E}/{\cal H}\r)}
\eeq
\end{widetext}
so that we have
\beq
\vert\chi_{E}(t)\vert^{2}
\simeq \l(\frac{1}{4 \pi \,\sqrt{k^2 + E}} \r)
\eeq
which is the amplitude of the wave function in flat space-time.
Evidently, substituting this wave function in the 
expression~(\ref{eq:mpsge}) for the modified spectrum will lead
to the required spectrum in flat space-time.

It should be emphasized here that we have arrived at the flat
space-time result by taking the limit of ${\cal H}\to 0$ 
{\it without}\/ imposing the condition of Hubble exit in the 
expression for the power spectrum in a de Sitter background. 
Physically, the limit of ${\cal H}\to 0$ is equivalent to 
sub-Hubble scales in exponential inflation and, therefore, 
this limit leads to the flat space-time result.
It should be noted that the condition of Hubble exit and the
limit ${\cal H}\to 0$ correspond to two extreme limits.
Hence, we could not have arrived at the flat space-time result 
by setting ${\cal H}=0$ in the power spectrum evaluated at Hubble 
exit in exponential inflation.


\end{document}